# Metal – Insulator transition in tin doped indium oxide (ITO) thin films: Quantum Correction to the electrical Conductivity

Deepak Kumar Kaushik, K Uday Kumar*, A. Subrahmanyam

Semiconductor Laboratory, Department of Physics, Indian Institute of Technology Madras, Chennai-600036, India

*kanapuram.udaykumar@gmail.com

## Abstract

Tin doped indium oxide (ITO) thin films are being used extensively as transparent conductors in several applications. In the present communication, we report the electrical transport in DC magnetron sputtered ITO thin films in low temperatures (25-300 K). The low temperature Hall effect and resistivity measurements reveal that the ITO thin films are moderately dis-ordered ($k_f l \sim 1$) and degenerate semiconductor. The transport of charge carriers in these disordered ITO thin films takes place via the de-localized states. The disorder effects lead to the well- known metal-insulator transition; this transition is observed at 110 K in ITO thin films. The metal-insulator behaviour is explained by the quantum correction to the conductivity (QCC); this approach is based on the quantum-mechanical interference effects in the disordered systems. The insulating behaviour is attributed to the combined effect of the weak localization and the electron-electron interactions.

# I. INTRODUCTION

The study of electrical transport in amorphous and disordered system has drawn the attention of several researchers[1] over the past three decades. It is well known that the classical approach of Boltzmann transport equation provides the information of carrier density, mobility, resistivity and the behavior of these quantities with temperature. In the low temperature regime there is a significant contribution of the quantum-mechanical interference effects[2], termed as quantum correction to conductivity (QCC). It is meditative that this QCC formulation explores the transport of carrier rigor in low temperature which depends on the disorder and the effective dimensionality of the system. This effect has been reported well to explain the electronic transport in many disordered systems, $SrRuO_3$[3], $LaNiO_3$[4], CdMnTe [5], InGaN[6], Ga:ZnO[7], etc.

Tin doped indium oxide (ITO) is an industry standard and most prominent among the transparent conducting oxide (TCO) thin films. Due to its benchmark properties like high transparency (>85% in visible range) and lower resistivity (1-10 × $10^{-4}$ ohm-cm), it is widely used in the optoelectronics applications like the flat panel displays, touch screens, solar cells, photo-catalysis etc., over the past five decades[8-13]. Extensive and exhaustive studies on the structural, optical and electrical properties have been carried out on ITO thin films. A fair amount of understanding on the material processing and properties has been achieved. However, the electrical behavior of ITO thin films especially at low temperatures and particularly the metal − insulator transitions still leaves a scope for further studies to investigate and understand the low temperature behavior of these ITO thin films. Advanced and strategic application of ITO coated on Kapton (polyimide) in satellites[14-16] (for thermal management and in electro-static discharge) is one practical situation where ITO experiences extremely low temperatures.

In the present investigation, tin doped indium oxide (ITO) thin films are prepared by reactive DC magnetron sputtering at room temperature. XRD and absorption measurements are carried out to confirm phase and the optical band gap of ITO thin film. Low temperature resistivity measurement shows metal-insulator transition in disordered ITO thin film. The metal – insulator transition is explained in the frame of quantum correction to the conductivity (QCC).

## II. EXPERIMENTAL METHOD

Among the several techniques available for the preparation of ITO thin films, reactive DC magnetron sputtering is most versatile and industrially adopted technique. In the present investigation, the ITO thin films have been prepared at room temperature (300 K) by reactive DC magnetron sputtering technique in a commercial sputtering system (Anelva, Japan). The preparation method very briefly is: the vacuum chamber was initially evacuated to $2 \times 10^{-5}$ mbar, sputter gas argon and reactive gas oxygen are allowed at 20 sccm and 8 sccm respectively (MKS Mass Flow Controller) in the chamber till the pressure reached $3.5 \times 10^{-3}$ mbar; the In:Sn (90:10 by weight) rectangular target ($380 \times 128$ mm$^2$) was energized to a power density of 0.14 W/cm$^2$ by an arc suppressed DC power supply (Advanced Energy, MDX series). These ITO thin films have been annealed at 673 K in vacuum for 60 minutes to enhance the optical transparency and electrical conductivity. The thickness of the ITO films was measured by surface-profilometer (Bruker). The ITO thin films are characterized for their crystalline nature by a Philips X'pert diffractometer (copper K$_\alpha$ radiation, $\lambda$ = 1.5418 Å). In order to study the optical properties of ITO thin films, transmission spectrum is measured in the wavelength range from 300 to 1100 nm with UV-Vis spectrophotometer (Cary-60). The electrical resistivity and Hall Effect measurements have been measured by an automated Lakeshore Hall Effect Measurement System (HMS 7604) in Van der Pauw configuration. The contacts were given by soldering of Indium dots on to a square shaped sample of area $10 \times 10$ mm$^2$. The ohmic nature of the contacts has been confirmed by the I-V measurements. The low temperature Hall Effect measurements are carried out in the temperature range 25 to 300 K, using Closed Cycle Helium Refrigerator (CCR) (Jani's Cryotronics) with lakeshore temperature controller (model 340). The maximum error in the carrier concentration and resistivity are within ± 2.5% and ±1% respectively.

## III. RESULTS AND DISCUSSION

The thickness of the ITO thin films is 300±10 nm approximately. Figure 1shows that the ITO thin film is polycrystalline. The film growth orientation is along (222) plane of cubic phase of ITO (JCPDS 01-089-4598). SnO$_2$ and metallic tin phases are absent which indicates that Sn replaced the In as a substitution dopant. The average crystallite size (D) is calculated using Scherrer's semi-empirical formula[17]

$$D = \frac{0.9\lambda}{\beta \cos \theta} \quad (1)$$

where $\lambda = 1.5418$ Å and $\beta = B - b$ (B = observed FWHM and b is broadening in peak due to instrument). The crystallite size is found to be 72 nm along (222) plane. The optical transparency of ITO film (Fig.2) is more than 85% in the visible range. The direct optical band gap E$_g$ is calculated by Tauc plot[18]-

$$(\alpha h\nu)^2 = A(h\nu - E_g) \quad (2)$$

and $\alpha$ is absorption co-efficient, given by

$$\alpha = \frac{1}{t} \ln \left[ \frac{(1-R)^2}{T} \right] \quad (3)$$

where $h$ is Planck's constant, $\nu$ is frequency of photon, $t$ is thickness of sample, $T$ is transmittance, $R$ is reflectance and $A$ is constant of proportionality. The optical band gap is calculated 3.87 eV which is higher than the non-degenerate ITO[19-21]. The higher value of optical band gap indicates that some states are filled in the conduction band which reveals that the ITO thin film prepared in the present investigation is degenerate semiconductor.

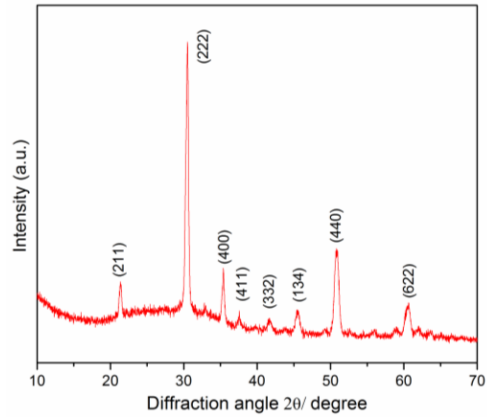

FIG.1. XRD pattern of reactive DC Magnetron sputtered and annealed tin doped indium oxide.

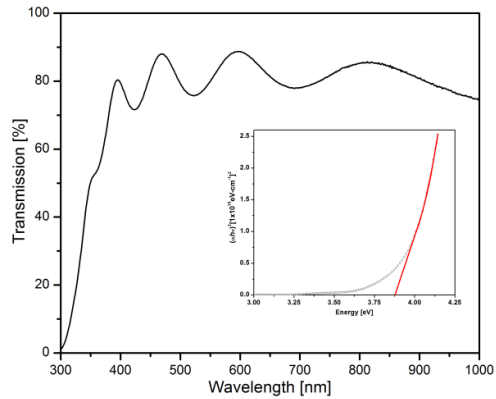

FIG. 2. Optical transmission spectrum of ITO thin film. The band gap is 3.87eV evaluated from Tauc's plot shown in the inset.

The room temperature electrical resistivity of ITO thin film is $5.5 \times 10^{-4}$ ohm-cm. The carrier density of electrons and hall mobility values are $7.2 \times 10^{-20}$ cm$^{-3}$ and 16.1cm$^2$/V/s respectively.

These values are similar to the earlier reported ITO thin films prepared by magnetron sputtering[22-24]. It is noted that the ITO film presented in the present report is a wide band gap and low resistivity semiconductor. Now we report temperature dependence of these electrical properties of ITO thin film.

The temperature dependent electrical resistivity of ITO thin film shows metal-insulator behavior at 110 K i.e., metallic behavior in the higher temperature range (>110 K) and insulating behavior below 110 K (Fig. 3). An understanding of this metal-insulator transition is the topic of the present communication and it is attempted through the frame work of the quantum correction to the conductivity model.

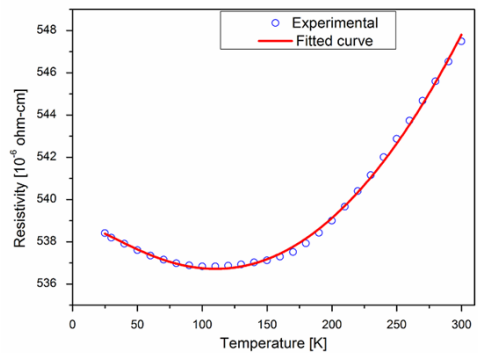

FIG. 3. Electrical Resistivity versus temperature plot of ITO thin films. Blue circles are experimental values and red solid line is theoretically fitted curve according to the equation (5).

The temperature dependent carrier concentration obtained from Hall Effect measurement system is given in Fig. 4 (a) shows that the carrier density is rather independent of the temperature; this observation indicates the degenerate nature of ITO[25]. The mobility in Fig.4 (b) is also found to be independent of the temperature which is the characteristics of a weakly disordered system.

It is well known that the Boltzmann transport equation of metals is explained on the basis of the quasi-classical approach. In the low temperature, the electrical resistivity may be expressed as:

$$\rho = \rho_0 + kT^n \qquad (4)$$

where $\rho_0$ is the resistivity due to the scattering of electron with impurities, lattice imperfections etc., k is constant, the exponent "n" will be 2 if the electron-electron scattering dominates. In the quasi-classical approach, the quantum interference effect was not included in $\rho_0$ (= $m^*/e^2\tau n$), $m^*$ being the effective mass. At low temperatures, the behavior of resistivity deviates significantly from the equation (4) thus warrants quantum corrections.

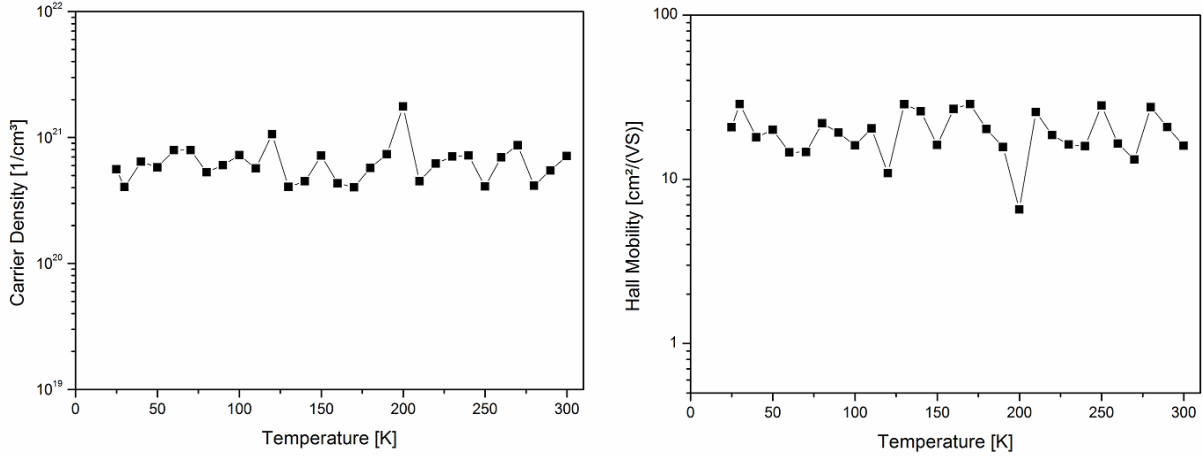

FIG. 4. (a) Carrier density versus temperature (b) Hall mobility versus temperature. Carrier concentration and Hall mobility behavior are independent of temperature (Solid line is drawn is a guide to the eye).

The basic theory of quantum correction to the conductivity (QCC) being reasonably well developed, the detailed study of QCC is given in reference 2. QCC is required at low temperatures, (i) when electron mean free path ($l$) is greater than that of the electron wavelength ($1/k_F$), $k_F l \gg 1$ (where $k_F$ is the Fermi wave vector), the quantum mechanical interference effect will be significant and (ii) where the probability of losing phase coherence of the electron is the least. It may be noted that the parameter $k_F l$ describes the disorder in the system. The quantum interference phenomenon dominates more at low temperatures where the inelastic collision frequency is so low that an electron undergoes elastic collisions without losing the phase coherence.

It may be mentioned that the quantum correction to the conductivity have been reported at higher temperatures for wide gap degenerate semiconductors like Indium Gallium Nitride (IGN)[6] at 150 K and ZnO[7] at 140 K.

**Metal – Insulator transition in ITO thin films:**

In the present experiment, the mean free path ($l=h/\rho e^2 n\lambda_e$) of the electron and Fermi (electron) wavelength $\lambda_e$ ($1/\lambda_e = k_F = (3\pi^2 n)^{1/3}$) are calculated. These values depend on the electrical resistivity ($\rho$) and the carrier (electron) density ($n$); the carrier density value is obtained from the Hall effec7t measurements. It is found that the mean free path ($l$) is greater than the wavelength of electron i.e., the condition: $k_F l > 1$ is satisfied as shown in figure 5. As the disorder parameter $k_F l$ is greater than 1 implies that the ITO thin film is moderately disordered system, in which the transport of charge carriers takes place via de-localized states[26]. Thus, the behavior of the resistivity with temperature can be explained by the QCC approach. The experimental observation of resistivity with temperature (Fig. 3) is fitted (red line) with the equation (5). The temperature co-efficient of resistivity (d$\rho$/dT) changes sign from positive (metal) to negative (insulator) with decreasing temperature at ~110K.

In this QCC approach to explain the electrical conductivity, the interference effect enhances the probability of finding the electron (the sum of the probabilities of the paths connected between two points) by twice to that of the classical probability[2]. In other words, the enhanced probability at self-crossing point reduces the probability of the electron at the point of observation, resulting in a corresponding reduction in the conductivity. Therefore, the quantum interference leads to a decrease in the conductivity. Thus the equation (4) is modified as[6, 27-29],

$$\rho = \frac{1}{\sigma_0 + mT^{1/2} + BT^{p/2}} + kT^2 \qquad (5)$$

where $\sigma_0$ is Boltzmann conductivity ($e^2\tau n/m^*$), $mT^{1/2}$ corresponds to the Columbic electron interactions (Altshuler-Aronov correction) and the term $BT^{p/2}$ describes the weak localization due to the self-interference of wave-functions backscattered on impurities; "p" depends on the nature of interaction (2 for electron-electron and 3 for electron-phonon interactions) [reference 11-13]. The parameters extracted on fitting the resistivity curve are given in Table I. Form equation 5, it may be noted that (i) the electron-electron interaction term ($mT^{1/2}$) increases with temperature and (ii) the contribution of weak localization term (BT) increases with decrease in temperature; the overall result is a decrease in the conductivity in non-metallic side of the metal-insulator transition. The negative value of "m" indicates the dominant contribution to the metallic behavior of ITO films. When the exponent p = 2 (equation 5 and Table 1), indicates that the electron-electron interaction is more dominating in the transport of carriers.

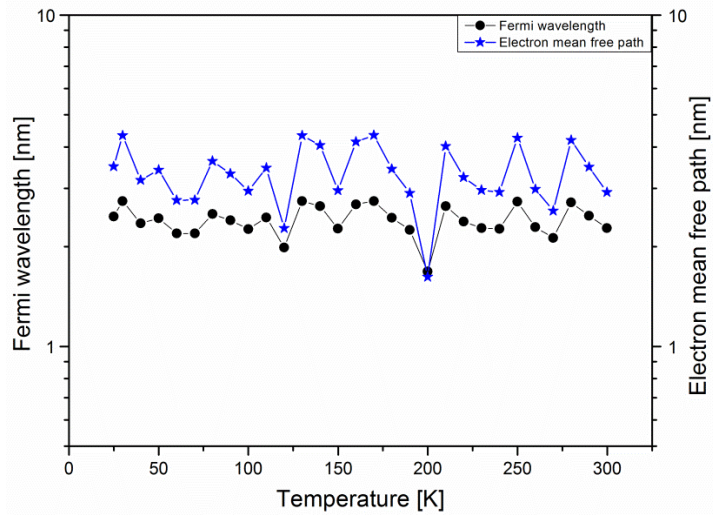

FIG. 5. Temperature variation of Fermi wavelength and electron mean free path.

| m (ohm-cm-K$^{1/2}$)$^{-1}$ | B (ohm-cm-K)$^{-1}$ | k (ohm-cm-K$^{-2}$) | p |
|---|---|---|---|
| -1.95 | 0.35 | 3.31x10$^{-10}$ | 2 |

Table I. Parameters used to fit equation (5)

The coefficient of weak localization term (BT) from equation 5, predicts the length scale, inelastic-collision length or Thouless length (L$_{Th}$), up to which quantum interference or localization effects are effective. The L$_{Th}$ can be evaluated from the equation[27]

$$B = \frac{e^2}{\hbar \pi^3 a} \quad (6)$$

where 'e' and '$\hbar$' are the electron charge and the Planks constant and 'a' is constant related to inelastic collision length as L$_{Th}$ ~ aT$^{-p/2}$. The value of L$_{Th}$ at 110 K is 2.04 nm and increases to 8.97 nm at 25 K which reveals that the quantum-mechanical interference effects are effective at large length scale in lower temperature. The lower values of L$_{Th}$ may be attributed to higher scattering rates.

## IV. CONCLUSION

Tin doped indium oxide thin film prepared by reactive DC magnetron sputtering is polycrystalline and wide band gap degenerate semiconductor. Tin doped indium oxide (ITO) thin films show a metal – insulator transition in electrical resistivity at 110 K. Hall effect measurements over the temperature range 25 – 300 K show a temperature independent carrier density and mobility. The metal – insulator transition is explained on the basis of quantum correction to the conductivity (QCC). QCC ascribes weak localization due to electron-electron scattering in the temperature range 25-110 K. In lower temperature range (<110 K) the quantum mechanical interference (weak localization) become more dominant.